\documentclass[12pt,a4paper,notoc]{JHEP}
\usepackage{amssymb,amsfonts}
\usepackage{latexsym}
\usepackage{amsmath}
\usepackage{fancybox}
\usepackage{frame}
\usepackage{slashed}
\usepackage{ulem}
\usepackage{color}
\usepackage{cite}
\usepackage{lscape}
\let\normalcolor\relax

\relax
\def\be{\begin{equation}}
\def\ee{\end{equation}}
\def\bea{\begin{eqnarray}}
\def\eea{\end{eqnarray}}

\newcommand{\bear}{\begin{eqnarray}}

\newcommand{\eear}{\end{eqnarray}}

\def\sq
\def\y{\psi}

\DeclareMathOperator{\Tr}{Tr}

\DeclareMathOperator{\arccosh}{arccosh}

\DeclareMathOperator{\cn}{cn}
\DeclareMathOperator{\nd}{nd}
\DeclareMathOperator{\cd}{cd}

\def\bR {\mathbb{R}}

\def\bZ {\mathbb{Z}}

\newcommand{\beq}{\begin{equation}}
\newcommand{\eeq}{\end{equation}}
\newcommand{\bal}{\begin{equation}\begin{aligned}}
\newcommand{\eal}{\end{aligned}\end{equation}}
\newcommand{\half}{\frac{1}{2}}

\newcommand{\eqn}[1]{(\ref{#1})}

\newcommand{\cN}{{\mathcal N}}

\newcommand{\cS}{{\mathcal S}}

\newcommand{\cI}{{\mathcal I}}

\title{\boldmath
The exact Schur index of $\cN=4$ SYM}

\author{ Jun Bourdier, Nadav Drukker and Jan Felix
\\
Department of Mathematics, King's College London,
\\
The Strand, WC2R 2LS, London, United-Kingdom
\\
\email{jun.bourdier@kcl.ac.uk},
\email{nadav.drukker@gmail.com},
\email{jan.felix@kcl.ac.uk}}

\preprint{}

\abstract{The Witten index counts the difference in the number of bosonic 
and fermionic states of a quantum mechanical system. The Schur index, which can 
be defined for theories with at least $\cN=2$ supersymmetry in four dimensions is 
a particular refinement of the index, dependent on one parameter $q$ serving as 
the fugacity for a particular set of charges which commute with the hamiltonian and 
some supersymmetry 
generators. This index has a known expression for all Lagrangian and some non-Lagrangian 
theories as a finite dimensional integral or a complicated infinite sum. 
In the case of $\cN=4$ SYM with gauge group $U(N)$ we rewrite 
this as the 
partition function of a gas of $N$ non interacting and translationally invariant 
fermions on a circle. 
This allows us to perform the integrals and write down 
explicit expressions for fixed $N$ as well as the exact all orders large $N$ expansion.}


\begin{document}



The superconformal index is a beautiful generalization of the Witten index \cite{Witten1982} to 
supersymmetric field theories. For four dimensional supersymmetric theories on $S^3\times\bR$ 
it is possible to define an index which depends on three fugacities, coupling 
to conserved charges which commute with the supersymmetry generator used 
to define the index.%
\footnote{In addition one can define fugacities for flavor symmetries.} 
Explicit expressions in terms of integrals of elliptic gamma 
functions exist for all Lagrangian theories
\cite{Romelsberger2006,Kinney2007,Dolan2009}.

The crucial property of an index is that it does not depend on continuous moduli 
of the theory. In the case of conformal $\cN=2$ theories there are many such moduli, 
the gauge couplings. These theories have a rich structure of conjectured S-dualities; seemingly 
different theories which are in fact the same under the appropriate mapping of 
couplings. Since the index should be the same for dual theories, it is possible 
to find expressions for the index of strongly interacting theories without a known 
Lagrangian (for a review, see \cite{Rastelli2014}). 
It was also shown that the expressions for the index satisfy the axioms 
of a topological field theory in 2d living on an auxiliary Riemann surface \cite{Gadde2010}.

A particularly simple version of the index of $\cN=2$ theories which depends only on a 
single fugacity is known as the Schur index \cite{Gadde2011,Gadde2013}. 
The contributions of both the vector and hyper 
multiplets can be expressed in terms of $q$-theta functions, rather than elliptic gamma 
functions \cite{Razamat2012}. 
The topological field theory is then the zero area limit of 
$q$-deformed two dimensional Yang-Mills and the 
form of the index of the strongly interacting theories are given by explicit infinite sums 
\cite{Gadde2011}. 

The purpose of this note is to write explicit expressions 
in terms of elementary functions for the Schur index of 
the simplest interacting field theory, namely $\cN=4$ 
SYM.%
\footnote{Other limits of fugacities of the index of $\cN=4$ SYM were studied 
in \cite{Spiridonov2012}.}

The integral representation of the Schur index of $\cN=4$ SYM with gauge 
group $U(N)$ is%
\footnote{For $SU(N)$ gauge group the prefactor should be rescaled by 
$q^{1/4} \eta^2(\tau/2) \eta^{-4} (\tau)$. In addition a delta function should be introduced so 
that sum $\alpha_i$ vanishes modulo $\pi$. The final answer is the same 
as for $U(N)$, up to the aforementioned prefactor. We thank Arash Arabi Ardehali for pointing out an omission in this factor. }
\beq
\label{4dvec}
\mathcal{I}(N)
=\frac{q^{-N^2/4}\eta^{3N}(\tau)}{N!\pi^N}
\int_0^\pi d^N \alpha \,
\frac{\prod_{i <j} \vartheta_1^2( \alpha_i-\alpha_j)}
{\prod_{i,j} \vartheta_4(\alpha_i-\alpha_j)}
\eeq
This expression is written in different notations than normally found in the 
literature; we use standard Jacobi theta functions and the Dedekind eta function 
rather than $q$-theta functions and $q$-Pochhammer symbols. Due to that, we also 
choose our $q$ to be the square root of that used more often. We also rescaled 
$\alpha$. Some simple definitions 
and relations of theta functions are given in Appendix~\ref{sec:identities}. 
We use the shorthand 
$\vartheta_i(z) = \vartheta_i(z,q)$ and $q \equiv e^{i\pi\tau}$. 

The product in \eqn{4dvec} can be rewritten using 
an elliptic determinant identity. For arbitrary 
$\alpha_i$ and $\alpha'_i$, one has \cite{Frobenius1879,Frobenius1882,Chudnovsky1993,Krattenthaler2005}
\beq
\frac{\prod_{ i < j} 
\vartheta_1 ( \alpha_i - \alpha_j ) 
\vartheta_1 (\alpha'_i - \alpha'_j ) }
{\prod_{i,j} \vartheta_4 (\alpha_i - \alpha'_j)} 
=\frac{ q^{-N^2/4}e^{-i N \sum_{i=1}^N (\alpha_i - \alpha'_i)}}
{\vartheta_3\big( \sum_{i=1}^N (\alpha_i - \alpha'_i)+ \pi \tau\frac{N}{2}\big)
\vartheta_3^{N-1} }
\det_{i,j}\left( \frac{ \vartheta_2 (\alpha_i - \alpha'_j)}
{\vartheta_4 (\alpha_i - \alpha'_j) } \right) . 
\eeq
Here we used the notation $\vartheta_i = \vartheta_i(0,q)$. 
Setting $\alpha'_i=\alpha_i$ simplifies the prefactor and 
we can also use the quasi-periodicity of 
$\vartheta_3$ to write
\beq
\label{delta}
\vartheta_3\big(\pi\tau N/2\big)
=q^{-N^2/4}\vartheta_3\Delta_N\,,
\qquad
\Delta_N 
= \begin{cases} 1\,, &\text{$N$ even} ,
\\
\frac{\vartheta_2}{\vartheta_3} \,, & \text{$N$ odd} \,.
\end{cases}
\eeq
Now we note that the ratio of Jacobi theta functions appearing in the determinant is in 
fact closely related to the elliptic function $\cn$ with modulus 
$k = \vartheta_2^2/\vartheta_3^2$
\beq
\frac{ \vartheta_2 ( z ) } { \vartheta_4 ( z ) }
= \frac{\vartheta_{2} }{\vartheta_4 } \cn (z \vartheta_{3}^2)\,.
\eeq
Finally using the relation 
$\vartheta_2\vartheta_3\vartheta_4
=2\eta^3(\tau)$ 
we can write the index \eqn{4dvec} as%
\footnote{The factor of $\vartheta_2^2/2\pi$ is included with the $\cn$ functions to simplify 
their Fourier expansion below.}
\beq
\label{SYMfermi}
\mathcal{I}(N) =\frac{q^{-N^2/4 }}{\Delta_N}\,Z(N)\,,
\qquad
Z(N)= \frac{1}{N!} \sum_{\sigma \in S_N} 
(-1)^\sigma \int_0^\pi
d^N\alpha \prod_{i=1}^N 
\frac{\vartheta_2^2}{2\pi}\cn \big( (\alpha_i - \alpha_{\sigma(i)} )\vartheta_3^2 \big)\,,
\eeq

Equation \eqn{SYMfermi} has the form of the partition function of $N$ 
\emph{free} fermions on a circle (see \emph{e.g.} \cite{Feynman1972}). 
The Fermi gas partition function is completely determined by the spectral traces 
\bal
\label{Zldef}
Z_\ell&= \Tr (\rho^\ell) = \int_0^\pi d \alpha_1 \dots d\alpha_\ell \, 
\rho\left( \alpha_1, \alpha_2 \right) \dots \rho \left( \alpha_\ell, \alpha_1 \right) \, ,
\\
\rho\big(\alpha,\alpha'\big)
&=\frac{\vartheta_2^2}{2\pi}\cn \big((\alpha - \alpha')\vartheta_3^2 \big)
=\frac{1}{\pi}\sum_{p\in\bZ}\frac{e^{i (2p-1)( \alpha - \alpha') }}{q^{p-\half}+q^{-(p-\half)}}\,,
\eal
where we used the Fourier expansion of the $\cn$ function.

Performing the integrals over $\alpha$ identifies the Fourier coefficients in the expansion of 
the different $\rho$'s giving an exceedingly simple result
\beq
\label{SUNZl}
Z_\ell
=\sum_{p\in\bZ}
\left(\frac{1}{q^{p-\half}+q^{-(p-\half)}}\right)^\ell.
\eeq

As we show in Appendix~\ref{sec:finiteN}, these infinite sums can be expressed as 
polynomials of complete elliptic integrals $K$ and $E$. One can then use the combinatorics 
of the conjugacy classes of the symmetric group $S_N$ with $m_\ell$ cycles of length 
$\ell$ to write $Z(N)$ \eqref{SYMfermi} as
\bal
\label{ZNZl}
Z (N) = {\sum_{\{m_\ell\}}}^\prime \prod_\ell \frac{ Z_\ell^{m_\ell} \, (-1)^{(\ell-1)m_\ell}} {m_\ell!\, l^{m_\ell} }\,,
\eal
where the prime denotes a sum over sets that satisfy $\sum_\ell \ell m_\ell = N $. 
Plugging \eqn{I_1234} into \eqn{ZNZl} and including the normalization in 
\eqn{SYMfermi}, we find for $N=1,2,3,4$
\bal
\label{finiteN}
\mathcal{I}(1) 
&= \frac{q^{-1/4}}{\pi} \sqrt{k} K \,,
\qquad\qquad
\mathcal{I}(2) = \frac{q^{-1}}{2\pi^2} K (K-E) \,,
\\
\mathcal{I}(3) 
&=\frac{q^{-9 /4}}{24\pi^3} \sqrt{k} K \big(12K(K- E)-4 (1+k^2) K^2+\pi ^2\big)\,,
\\
\mathcal{I}(4) 
&= \frac{q^{-4}}{24\pi^4} K\big( 3K(K-E)^2 - 2 k^2K^3+\pi ^2 (K-E) \big) \,.
\\
\eal
It is easy to generate explicit expressions for larger $N$. 
The powers of $q$ in these expressions 
as well as $\sqrt k$ for odd $N$ come from the normalization prefactor in \eqn{SYMfermi}. 
We see that by removing this power of $q$, the resulting expressions are 
polynomials of complete elliptic integrals and the elliptic modulus and have nice 
modular properties. We comment on that further below.

An alternative approach is to consider the grand canonical partition function
with fugacity $\kappa$
\beq
\label{grandpartition}
\Xi(\kappa) = 1 + \sum_{N=1}^\infty Z(N) \kappa^N \,.
\eeq
This is a Fredholm determinant of a very simple form
\beq
\label{SUNXi}
\Xi(\kappa) = \exp\left(-\sum_{\ell= 1}^\infty \frac{ (-\kappa) ^\ell}{ \ell}Z_\ell\right)
= \prod_{p \in \mathbb{Z}} \left( 1 + \frac{\kappa}{q^{p-\half}+q^{-(p-\half)}} \right)
\eeq
This product turns out to be expressible in terms of theta functions 
(see Appendix~\ref{sec:identities})
\bal
\label{xi-exact}
\Xi(\kappa) 
&= \prod_{p=1}^\infty \left( \frac{1 + q^{2p-1} + \kappa q^{p-\half} }
{1 + q^{2p-1} }\right)^2 
= \frac{\left(\prod_{p=1}^\infty(1-q^p)(1 + q^{2p-1} + \kappa q^{p-\half})\right)^2 }
{\prod_{p=1}^\infty(1-q^p)^2(1 + q^{2p-1})^2}
\\&= \frac{1}{\vartheta_3\vartheta_4}
\vartheta_3^2 \Big( \frac{1}{2} \arccos \frac{\kappa}{2} , q^{1/2}\Big)
=\frac{1}{\vartheta_4}
\left(\vartheta_3\left( \arccos\frac{\kappa}{2}\right)
+\frac{\vartheta_2}{\vartheta_3}\,\vartheta_2 \left(\arccos \frac{\kappa}{2}\right)\right).
\eal

Recall that the index has the extra factor of $1/\Delta_N$ \eqn{delta} compared 
with the free fermion partition function \eqn{SYMfermi}. This factor 
distinguishes even and odd $N$, so it is natural 
to split $\Xi$ into its even and odd parts. In fact, since $\vartheta_3(z)$ 
is even and $\vartheta_2(z)$ is odd under $z\to\pi-z$, the 
two terms in the last line of \eqn{xi-exact} give the decomposition into
\beq
\label{Xi-pm}
\Xi_\pm
\equiv\frac{1}{2}\Big(\Xi(\kappa)\pm\Xi(-\kappa)\Big)
\eeq
It is then possible to define the \emph{grand index} $\hat{\Xi}$, 
where the $\Xi_-$ part is divided by the value of $\Delta_N$ for odd $N$
\beq
\label{xi-hat}
\hat{\Xi}(\kappa) 
\equiv 1 + \sum_{N=1}^\infty \cI(N) q^{N^2/4} \kappa^N
= \frac{1}{\theta_4} \left[ \vartheta_3\left(\arccos \frac{\kappa}{2}\right) 
+ \vartheta_2 \left(\arccos \frac{\kappa}{2}\right) \right].
\eeq
By expanding the last expression in powers of $\kappa$, it is very easy 
to obtain explicit formulae for $\cI(N)$ for finite $N$. Using relations between derivatives 
of theta functions and elliptic integrals, one recovers \eqn{finiteN}.

We can also obtain a closed formula for the coefficients of the large $N$
expansion of $\cI(N)$ by expanding \eqn{xi-hat} at large $\kappa=e^\mu$ and 
using the integral transform
\beq
\label{ZNJmu}
\cI(N) = \frac{q^{-N^2/4}}{2\pi i } \int_{ - i \pi }^{i \pi }d\mu\, \hat\Xi(e^\mu) e^{ - \mu N} \, ,
\eeq
Given the large $\kappa$ expansion of $\arccos$
\beq
\arccos\frac{e^\mu}{2}
=i\arccosh \frac{e^\mu}{2}
=i\mu+i\log \frac{1+\sqrt{1-4 e^{-2 \mu }}}{2}\,,
\eeq
we find
\beq
e^{-\mu N}
\hat\Xi(e^\mu)
=\frac{e^{-\mu N}}{\vartheta_4}
\,\text{\Large:}e^{\log \frac{1+\sqrt{1-4 e^{-2 \mu }}}{2}\partial_\mu}\text{\Large:}\,
\big( \vartheta_3 ( i\mu) + \vartheta_2 ( i\mu) \big)\,,
\eeq
where the exponent is normal ordered. Expanding in powers of 
$e^{-2 \mu } $ gives
\beq
e^{-\mu N}
\text{\Large:} e^{\log \frac{1+\sqrt{1-4 e^{-2 \mu }}}{2}\partial_\mu}\text{\Large:} 
=e^{-\mu N}+\sum_{n=1}^\infty \frac{(-1)^n}{n!} e^{-(N+2 n) \mu} 
\partial_\mu \prod_{k=1}^{n-1} (\partial_\mu - n - k ) \,.
\eeq

The index \eqn{ZNJmu} can now be easily evaluate through integration by parts. 
For any integer $l\geq 0$, we have
\beq
\frac{1}{2 \pi i } \int_{-i \pi}^{i \pi} d\mu\, e^{- \mu \left(N + 2n\right)} 
\partial_\mu^l \vartheta_3( i \mu) 
= \frac{( N+2n)^l}{2 \pi i}\int_{-i \pi}^{i \pi} d\mu\,
e^{- \mu (N + 2n)} \vartheta_3( i \mu)
\eeq
and similarly for $\vartheta_2( i \mu)$. Using the Fourier expansion
\beq
\frac{1}{2\pi i} \int_{-i \pi}^{i \pi} d\mu\,e^{- \mu \left(N+2n\right)} 
\big( \vartheta_3( i \mu) + \vartheta_2( i \mu)\big) 
= q^{ \frac{\left(N+2n\right)^2}{4}}\,,
\eeq
we finally obtain
\beq
\label{largeN}
\cI(N)
=\frac{1 }{\vartheta_4}
\sum_{n=0}^\infty (-1)^n \left[\binom{N+n}{N}+\binom{N+n-1}{N}\right]q^{nN+n^2}\,.
\eeq
This sum is convergent for $|q|<1$ and it is easy to check that for $N=1,2,3,4$ it 
is exactly the $q$-expansion of the finite $N$ expressions in \eqn{finiteN}.

Equations \eqn{largeN}, \eqn{xi-hat}, \eqn{xi-exact} and \eqn{finiteN} 
(with the algorithm in Appendix~\ref{sec:finiteN}) are the main results of this paper. 

The large $N$ expansion \eqn{largeN} should have an interpretation in string theory, but it 
is a long enduring puzzle to get a match. The leading contribution 
to the Schur index with $n=0$ is just $1/\vartheta_4$, which is $N$ independent, unlike the 
supergravity action, which should scale with $N^2$. 
This was already observed for the large $N$ limit of the most general index of $\cN=4$ SYM 
in \cite{Kinney2007}. Indeed, the correct quantity to match should be the partition function 
on $S^3\times S^1$, which is related to our definition of the index by the Casimir factor, which 
does scale as $N^2$ \cite{Assel2014,Assel2015}. It still does not match the gravity 
calculation, possibly due to missing counterterms of supersymmetric holographic renormalization.

Still one could hope 
that the terms with $n>1$ correspond to some instantons in string theory. Identifying 
$\log1/q$ with the radius of the compact $S^1$ (relative to the radius of $S^3$), it 
seems to match the action of $n$ D3-branes plus $n^2$ excitations between the 
branes. Natural candidates would be $1/8$ BPS giant gravitons 
\cite{Mikhailov:2000ya,Biswas:2006tj,Mandal:2006tk}, which 
should play a similar role to the string and membrane instantons in the 
case of ABJM \cite{Aharony2008,Drukker:2010nc}.

The grand canonical partition function $\Xi$ \eqn{xi-exact} and the grand index 
$\hat \Xi$ \eqn{xi-hat} bear a very strong resemblance to the grand canonical partition
function 
of $\cN = 8$ superconformal Chern-Simons-matter theories, 
which was expressed in \cite{Codesido2015} also 
in terms of a Jacobi theta function. In the $3$-dimensional case, the classical limit 
of the grand potential for a wide class of theories
contains a cubic term in $\mu$ which leads to the universal Airy 
function behavior for the partition function \cite{Marino2012}. 
The analog in $4$ dimensions is a 
quadratic term in $\mu$ (related the the structure of the Fermi surface on the circle). 
The analog of the Airy function is then a $q^{kN^2}$ scaling of the partition function of the 
fermions, here with $k=1/4$. But this behavior is not evident in the 
final answer, due to the overall prefactor in the definition \eqn{4dvec}, \eqn{SYMfermi}. 
If the $\mu^2$ scaling is universal, it would be natural to try to reproduce it 
from a quantum supergravity calculation, along the lines of 
\cite{Dabholkar2014}, though an understanding of classical supergravity 
is required first.

Comparing to the Fermi-gas formulation of 3d theories \cite{Marino2012}, 
the expressions we find are even simpler, since the density 
$\rho$ \eqn{Zldef} is a function of difference form, which means that the Hamiltonian 
$H=-\log\rho$ is a function only of the momentum. So the free fermions 
have a complicated kinetic term, but no potential and no interactions. Indeed the 
exact spectrum is pairwise degenerate with
\beq
E_p=\log\Big(q^{p-\half}+q^{-p+\half}\Big)
=-\left(p-\half\right)\log q-\sum_{l=1}^\infty\frac{(-1)^l}{l}q^{l(2p-1)}
\,,\qquad
p\geq1\,.
\eeq
Identifying $\hbar=-i\pi\tau$, this is the spectrum of the harmonic oscillator with 
exponentially small corrections at large $\hbar$. 
In the 3d case there is a connection between the grand canonical partition 
functions and the spectrum of Schr\"odinger equations on the line related 
to topological string theory \cite{Huang:2014eha ,Grassi2014}. One can easily enrich 
our model such that it would calculate non-trivial spectra on the circle. 
A natural modification 
is to include extra fundamental hypermultiplets. While the resulting matrix 
model looks interesting, the 4d field theory is not asymptotically free, so it is not 
obvious at all what this would be capturing.

It was advocated in \cite{Razamat2012} that the index would have nice modular 
properties once rescaled by $q^c$, where $c$ is the central charge. This rescaling is exactly 
that mentioned after \eqn{finiteN} and is also the one appearing in our definition of the grand 
index $\hat\Xi$ \eqn{xi-hat}. Indeed for finite $N$ we find elliptic integrals and at fixed 
$\mu$ we get theta 
functions of nome $q$, all of which have nice modular properties. 
It was also proposed in \cite{Razamat2012} to modify the index by changing some 
charge assignments, which amounts in our case to introducing a mass $m=\pi/2$ for 
the adjoint hypermultiplet. It is easy to repeat our calculation for this particular mass 
and the result for the grand index is equal to our $\Xi_+(i e^\mu,-q)$ \eqn{Xi-pm}. 
The modified index therefore 
vanishes for all odd $N$, and is essentially the same as the usual index for even $N$. 

It would also be interesting to reproduce our expression from the topological field theory 
formulation as a sum over $U(N)$ representations \cite{Gadde2010, Gadde2013}. 
Given the explicit relation in this case to $q$-deformed 2d Yang-Mills \cite{Gadde2011}, 
it should also be possible to reproduce our results from an analysis similar to
\cite{Minahan1994}. 
The Schur index is also conjectured 
to be equal to the torus partition function of
certain 2d chiral algebras \cite{Beem2015}.
For $\cN=4$ SYM the 
conjectured chiral theory has super $\mathcal{W}$-algebra 
symmetry and our expressions should give the full 
partition function of this theory. In addition there is a relation between the index and integrable 
models \cite{Yamazaki2014}, which should also yield the same result.

One can also apply such ideas 
to other theories: Does the grand index take a simple form for all class-$\cS$ theories?
The most obvious generalization is
necklace quivers, where a lot of the structure persists.  
It is also possible to include 
fugacities for the flavor charges, which in the case of $\cN=4$ is analogous to considering 
$\cN=2^*$. 
Other generalizations are the inclusion of line operators, for which the 
matrix models were computed in the Schur limit in \cite{Gang2012}. 
It would also be interesting to see whether it is possible to generalize our analysis 
beyond the Schur limit and/or to the lens space index \cite{Benini2012}.
We hope to report on some of these issues 
in the near future.

\section*{Acknowledgements}
We would like to thank Stefano Cremonesi, Jo\~ao Gomes, Jaume Gomis, Ivan Kostov, 
Joe Minahan, Sanefumi Moriyama, Kostya Zarembo 
and the KCL journal club for enlightning discussions.
N.D. Would like to thank the hospitality of the IFT, Madrid, during the course of this work. 
The research of J.B. has received funding from the People Programme 
(Marie Curie Actions) of the European 
Union's Seventh Framework Programme FP7/2007-2013/ under REA 
Grant Agreement No 317089 (GATIS). 
The research of N.D. is underwritten by an STFC advanced fellowship.
The research of J.F. is funded by an STFC studentship ST/K502066/1.

\appendix

\section{Theta functions}
\label{sec:identities}

In the literature on the superconformal index, the expressions are mostly written 
in terms of $q$-theta functions and $q$-Pochhammer symbols. 
Those are related to the Jacobi theta functions and Dedekind eta function by
\bal
\label{qtojacobi}
\theta ( e^{2 iz} , q^2 ) 
= \frac{-ie^{iz}\,\vartheta_1 (z, q )}{q^{1/6}\eta(\tau)} \, ,
\qquad
\left(q^2 ; q^2\right)_\infty 
&= \prod_{k =1}^\infty \left( 1 -q^{2k} \right) 
=q^{-1/12}\eta(\tau)\,.
\eal
where the (quasi) period $\tau$ is related to the nome $q$ by $q= e^{i \pi \tau} $. 

The Jacobi theta function $\vartheta_3(z,q)$ is given by the series 
and product representations
\beq
\label{theta3}
\vartheta_3(z ,q) = \sum_{n=-\infty}^\infty q^{n^2} e^{2 i n z }
= \prod_{k=1}^\infty \left( 1 - q^{2k} \right) 
\left( 1 +2 q^{2k-1} \cos(2z) + q^{4k - 2} \right),
\eeq
in terms of which the auxiliary theta functions are
\bal
\label{auxtheta}
\vartheta_1 (z ,q ) &= i q^{1/4} e^{ - i z } \vartheta_3 ( z - \tfrac{1}{2} \pi \tau - \tfrac{1}{2} \pi ,q ) 
\\\vartheta_2 (z ,q) & = q^{1/4} e^{ -i z} \vartheta_3 ( z - \tfrac{1}{2} \pi \tau ,q ) 
\\\vartheta_4 (z,q ) &= \vartheta_3 ( z - \tfrac{1}{2} \pi ,q ) \,.
\eal
We also need Watson's identity 
(eq. 20.7(v) in \cite{Olver2010})
\beq
\label{Watson}
\vartheta_3 \left( z, q^{1/2} \right) \vartheta_3 \left( w, q^{1/2} \right)
= \vartheta_3 ( z + w, q) \vartheta_3 (z-w, q) 
+ \vartheta_2 ( z + w, q) \vartheta_2 (z-w, q)
\eeq

\section{Finite $N$ expressions}
\label{sec:finiteN}

The quantities $Z_\ell$ \eqn{SUNZl} can be evaluate 
by the procedure given in \cite{Zucker1979} which we briefly summarise 
(correcting some small typos). 
For even $\ell=2s$ one has
\beq
Z_{2s} = 2\frac{(-1)^{s+1}}{(2s-1)!} \sum_{m=1}^{s-1} \alpha_m(s) D_{2s-2m-1}\, ,
\eeq
where $\alpha_{m} (s)$ and $D_s$ are generated by 
\beq
\sum_{m=0}^{s-1}\alpha_m(s)t^m=\prod_{j=1}^{s-1}(1-j^2t)\,,
\quad
\sum_{s=0}^\infty D_{2s+1} \frac{(-4t^2)^{s}}{(2s)!} = \frac{EK}{2 \pi^2}
- \frac{K^2}{2 \pi^2} (1-k^2) \nd^2 \left(\frac{2 K t}{\pi}\right) ,
\eeq
where $K \equiv K(k^2)$ and $E \equiv E(k^2)$ are complete elliptic integrals of the 
first and second kind respectively, $\nd$ is an elliptic function (likewise $\cd$ below) and 
the elliptic modulus is $k = \vartheta_2^2/\vartheta_3^2$.

Similarly for odd $\ell=2s+1$
\beq
Z_{2s+1} = \frac{(-1)^s}{2^{2s-1} (2s)!} \sum_{m=0}^s \tilde \alpha_{m}(s) J_{2s-2m} \, ,
\eeq
where
\beq
\label{alphatilde}
\sum_{m=0}^s\tilde \alpha_{m}(s)t^m=\prod_{j=1}^{s}\big(1-(2j-1)^2t\big)\,,
\qquad
\sum_{s=0}^\infty  J_{2s} \frac{(-t^2)^s}{(2s)!} 
= \frac{ K k }{2 \pi} \cd \left(\frac{2 K t}{\pi}\right).
\eeq

We find for $N=1,2,3,4$
\bal
\label{I_1234}
Z_1&=\frac{kK}{\pi}\,,
\qquad&
Z_2&=\frac{KE-(1-k^2)K^2}{\pi^2}\,,
\\
Z_3&=\frac{Z_1}{8}-\frac{(1-k^2)kK^3}{2 \pi^3}\,,
\qquad&
Z_4&=\frac{Z_2}{6}-\frac{(1-k^2)k^2 K^4}{3\pi^4}\,.
\eal
Plugging these expressions into \eqn{ZNZl} and including the normalization in 
\eqn{SYMfermi} gives \eqn{finiteN}.
This algorithm can be easily pushed to larger values of $N$ 
and is easy to implement in Mathematica.

\addtolength{\parskip}{-.5mm}

\bibliographystyle{utphys2}

\bibliography{References}
\addtolength{\parskip}{.5mm}

\end{document}